\documentclass[11pt]{article}

\usepackage[]{inputenc}
\usepackage[T1]{fontenc}
\usepackage[hidelinks]{hyperref} 
\usepackage{mathrsfs}
\usepackage{color}
\usepackage{authblk}

\usepackage{fullpage}

\usepackage{amsmath}
\usepackage{amssymb}
\usepackage[toc,page]{appendix}

\usepackage{amsthm}
\usepackage{float}
\usepackage{placeins}
\usepackage{wrapfig}
\usepackage{caption}
\usepackage{graphicx}
\usepackage{pgfplots}
\usepackage{subcaption}
\usepackage{wasysym}
\usepackage{mathtools}

\newcommand{\sgn}[0]{\operatorname{sgn}}

\title{The merger of a black hole with a cosmological horizon}

\date{\today}

\pgfplotsset{compat=1.18}

\begin{document}

\author[1]{Maxime Gadioux\thanks{mjhg2@cam.ac.uk}}
%\author[1]{Harvey Reall\thanks{hsr1000@cam.ac.uk}}
\author[1]{Hangzhi Wang\thanks{hw586@cam.ac.uk}}
\affil[1]{\small Department of Applied Mathematics and Theoretical Physics, University of Cambridge, Wilberforce Road, Cambridge CB3 0WA, United Kingdom}

\maketitle

\begin{abstract}
    \noindent In recent years there have been many studies on exactly solvable black hole mergers, based on a model by Emparan and Mart\'inez where the mass of one black hole is blown up to infinity \cite{Emparan:2016ylg}. Here we replace the large black hole by a cosmological horizon, and study how it merges with a black hole in the Schwarzschild-de Sitter spacetime by considering an observer positioned at future null infinity. We describe the geometry of the horizon over time, including the role that caustics play in the merger process, and also examine the growth of the horizon area. We argue that in the limit of zero cosmological constant, the system reduces to the Emparan-Mart\'inez Schwarzschild merger. This allows us to regularise the increase in the area during the merger, which otherwise diverges.
\end{abstract}

\newpage

\section{Introduction}

With the advent of gravitational wave observations, black hole mergers have taken a prominent position in gravitational research. Unfortunately, the defining characteristic of black holes, their event horizon, is generally very difficult to identify due to its teleological nature: one must know the complete future evolution of the spacetime in order to determine it. However, Emparan and Mart\'inez showed that the problem simplifies considerably if one takes the limit of infinite mass ratio \cite{Emparan:2016ylg}, where the small black hole's mass $M$ is fixed and the large black hole is blown up to infinite size. In an extreme mass ratio merger, the curvature of the large black hole becomes negligible over lengthscales similar to the size of the small black hole, and in fact vanishes in the infinite mass ratio limit. Hence, the spacetime is described exactly by the metric of an isolated small black hole, e.g.~Schwarzschild. However, the event horizon is not the usual Schwarzschild horizon at areal radius $2M$, but a different null hypersurface that asymptotes to a null plane at future infinity, representing the final merged, infinitely-large black hole.

This idea has been employed in numerous papers since the original article, replacing the small Schwarzschild black hole with a Kerr black hole \cite{Emparan:2017vyp,Gadioux:2024tlm}, a neutron star \cite{Emparan:2020uvt}, a Reissner-Nordstr\"om black hole \cite{Pina:2022dye}, a wormhole \cite{Dias:2023pdx} and a spherically-symmetric black hole in Einsteinian cubic gravity \cite{Dias:2024wib}. Apart from Kerr, the remaining cases have a spherically-symmetric small object. In this configuration (and also for axisymmetric Kerr mergers), the event horizon is smooth everywhere, except on a line of caustic points. New generators enter the horizon along this line; in fact, owing to the axisymmetry of the event horizon, infinitely-many generators enter at each point on the caustic line. In the terminology of Refs.~\cite{Gadioux:2023pmw,Gadioux:2024tlm}, the caustic line forms the crease set, which generically plays a key role in the merger of black holes. The non-axisymmetric Kerr case provides a more realistic picture of an extreme mass ratio merger, as the event horizon no longer has any symmetry (except for a reflection symmetry for mergers in the equatorial plane---so-called orthogonal mergers). In these scenarios, the crease set spreads out into a $2$-dimensional surface of crease points bounded by caustic lines. Two generators enter the horizon at each crease point, whereas only one enters at the new caustic points. Ref.~\cite{Gadioux:2024tlm} showed in detail how the model of Emparan and Mart\'inez applied to the Kerr black hole agrees with generic results of how black holes in General Relativity merge \cite{Gadioux:2023pmw}.

While the exact description of a merger in the infinite mass ratio is a very valuable model, the unbounded size of the large black hole makes certain results difficult to interpret. For example, one would like to study the increase in horizon area, but a simple calculation shows that, to leading order, it is proportional to the product of the masses of the black holes, which blows up as the large black hole grows to infinite mass \cite{Emparan:2016ylg}. One can of course examine the area increase from the generators corresponding to the small black hole, but this leaves one half of the story incomplete.

To obtain the full picture, one would need an exact description of a merger that does not involve setting a parameter to infinity. One example is the merger of a black hole horizon with a cosmological horizon in Schwarzschild-de Sitter spacetime, as seen from an observer that is not in the rest frame of the black hole, i.e.~an observer who has a different cosmological horizon from that of the black hole. In this setting, there is one black hole, of mass $M$, and the only other parameter is the cosmological constant $\Lambda > 0$. 

In this paper, we study how the merger occurs, identify the caustic points on the horizon, and determine how the area changes as a function of time. The caustic points are crucial features: it is at such a point that the black hole and cosmological horizon first touch, and it is also there that the horizon develops conical singularities. We find that the increase in area originates mostly from generators that enter the horizon through caustic points, even when $\Lambda$ is very small with respect to $M$, and therefore the cosmological horizon is very large compared to the black hole. Moreover, we argue, using the concept of limits of spacetimes introduced by Geroch \cite{Geroch:1969ca}, that the $\Lambda\to 0$ limit of our system reduces to the infinite mass ratio merger of two Schwarzschild black holes, as originally studied by Emparan and Mart\'inez \cite{Emparan:2016ylg}. This offers a possible method of regularisation of divergent quantities in the latter system, such as the area growth of the large black hole. It seems likely that a similar procedure should also hold for rotating black holes. Finally, we show that our system satisfies a universal prediction made in \cite{Gadioux:2023pmw} regarding the scaling of the opening angles of the conical singularities in an axisymmetric merger.

The paper is organised as follows. In Section \ref{sec:event_horizon} we describe how we determine the event horizon, including solving the null geodesic equations and identifying the caustic points. We outline our results in Section \ref{sec:results}, beginning with describing the qualitative features of the merger, then studying the growth in area and the $\Lambda\to0$ limit, and finally examining the conical singularities on the horizon. We provide a brief summary and some further comments in Section \ref{sec:conclusion}. Throughout we will use units such that $G=c=1$. This implies that $\Lambda$ has dimensions of inverse mass squared, i.e.~$M\sqrt{\Lambda}$ is dimensionless. Our signature is $(-1,1,1,1)$.

\section{Determining the event horizon}\label{sec:event_horizon}

In this article, we study the merger of a Schwarzschild-de Sitter black hole with the cosmological horizon of an observer that is not at rest with the black hole. In the observer's frame, at early times (for a suitable notion of time), there is a non-rotating black hole whose horizon is disconnected from the cosmological horizon. The black hole moves away from the observer and towards the cosmological horizon, until the two horizons meet and merge. At late times, the black hole has disappeared behind the cosmological horizon, and the observer sees the same structure that they would find in a pure de Sitter universe.

\begin{figure}
    \centering
    \includegraphics[width=0.5\textwidth]{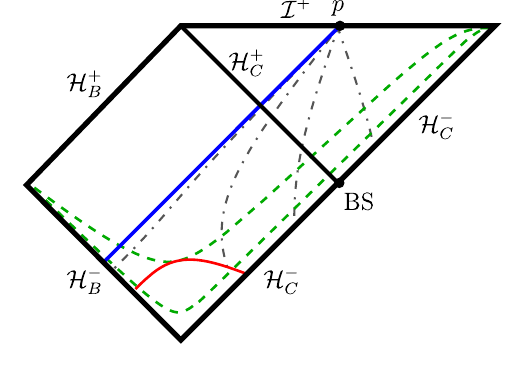}
    \caption{Penrose diagram of the relevant regions of Schwarzschild-de Sitter spacetime for our construction, with $\Lambda=0.1$, $M=0.5$. We choose a point $p$ on $\mathcal{I}^+$ to represent the accelerating observer. The thick solid blue line from $p$ is the radial null geodesic with $u\equiv0$ and parameters $q=0$, $E>0$. The solid red line extending from $\mathcal{H}_B^-$ to $\mathcal{H}_C^-$ is the (spacelike) caustic line. Some generators of the horizon $\mathcal{H}^+$ are shown in dash-dotted grey; these appear as timelike curves because the geodesics are not radial. Two constant-$T$ surfaces are shown in dashed green: one before merger, which intersects the caustic line twice, and another at the time of merger. The bifurcation sphere on the cosmological horizon $\mathcal{H}^\pm_C$ is denoted ``BS''.}
    \label{fig:penrose}
\end{figure}

The above was formulated on a more mathematical footing in \cite{Emparan:2016ylg,Emparan:2017vyp,Gadioux:2024tlm}, which we briefly repeat and adapt to the Schwarzschild-de Sitter case here. The event horizon of the merger is a null hypersurface that, near future null infinity $\mathcal{I}^+$, approaches an outgoing spherical null hypersurface (i.e.~of constant retarded time), representing the cosmological horizon of the observer after the merger has occurred. The observer, who is accelerating in the black hole's frame of reference, has a worldline that intersects $\mathcal{I}^+$ at a point $p$, and the horizon $\mathcal{H}^+$ as seen by the observer is the boundary of the causal past of $p$, $\mathcal{H}^+=\dot{J}^-(p)$. This is illustrated in the Penrose diagram in Figure \ref{fig:penrose}. Of course, the Penrose diagram still features the black and white hole horizons $\mathcal{H}^\pm_B$ and the past and future cosmological horizons $\mathcal{H}^\pm_C$ of the usual Schwarzschild-de Sitter spacetime, but these are not directly relevant for the actual horizon of the merger. We emphasise that $\mathcal{H}_C^\pm$ is \emph{not} the cosmological horizon of the observer, but instead that of the black hole.

A generator of $\mathcal{H}^+$ may have a past endpoint, which is the point at which it enters the horizon (the null geodesic may be extended to the past beyond the past endpoint, but the extended segment is outside the horizon). Starting from $p$ and evolving backwards in time, the past endpoint along a specific generator $\gamma$ is the first point of intersection between $\gamma$ and any other generator of $\mathcal{H}^+$.\footnote{The point of intersection will often be a past endpoint of multiple generators, but may sometimes be the past endpoint of only one generator \cite{Gadioux:2023pmw}, such as for the caustic points on the horizon of a merger of Kerr black holes \cite{Gadioux:2024tlm}.} Owing to the symmetry of the spacetime under consideration, the past endpoints in our case are all caustic points. A generator evolved backwards from $p$ may encounter a caustic point, or it may cross the white hole horizon $\mathcal{H}^-_B$ or the past cosmological horizon $\mathcal{H}^-_C$ before doing so. Thus, one can categorise all generators into three groups according to their origin. We shall call them \emph{caustic generators}, \emph{black hole generators}, and \emph{cosmological horizon generators}, respectively. We foliate our spacetime with spacelike surfaces extending from the bifurcation sphere of the black hole (left vertex in Figure \ref{fig:penrose}) to the intersection of $\mathcal{H}_C^-$ and $\mathcal{I}^+$ (top-right vertex). We choose the foliation such that the time function $T$ is the Killing time measured from one of the spacelike surfaces. Then the infinite past is the union of the white hole horizon and the past cosmological horizon, $\mathcal{H}_B^- \cup \mathcal{H}_C^-$, while the infinite future is $\mathcal{H}_B^+ \cup \mathcal{I}^+$. Our time function will be precisely defined later. The cross-section $\mathcal{H}_B^-\cap\mathcal{H}^+$ gives the black hole in the infinite past, while $\mathcal{H}_C^-\cap\mathcal{H}^+$ describes the observer's cosmological horizon, also in the infinite past. These surfaces are disconnected.

To make the discussion more concrete, let us introduce spherical polar coordinates centred on the black hole, $(t,r,\theta,\phi)$. The metric reads
\begin{equation}\label{eq:metric_t}
    ds^2=-f(r)dt^2+f(r)^{-1}dr^2+r^2(d\theta^2+\sin^2\theta\, d\phi^2),
\end{equation}
where $f(r)=1-2M/r-\Lambda r^2/3$, $M$ is the black hole mass and $\Lambda$ is the cosmological constant. The function $f(r)$ has two distinct positive roots, $0<r_B<r_C$, whenever $0<\Lambda<1/9M^2$ and $M>0$. These satisfy $2M<r_B<3M<1/\sqrt{\Lambda}<r_C<3/\sqrt{\Lambda}$. The smaller root, $r_B$, gives the radius of $\mathcal{H}_B^\pm$, whereas $r_C$ is the radius of $\mathcal{H}_C^\pm$. Properties of depressed cubic functions ensure that the third root of $f(r)$ is $-r_B-r_C<0$, and we can therefore write
\begin{equation}
    f(r) = \frac{\Lambda}{3}\frac{(r-r_B)(r_C-r)(r+r_B+r_C)}{r}.
\end{equation}
Clearly, $f(r)$ is positive between $r_B$ and $r_C$ and negative otherwise (restricting to $r\geq 0)$.

As we are working with outgoing null geodesics, it is useful to introduce the retarded time $u\equiv t-r^\star$, where $r^\star$ is the tortoise coordinate defined by $dr^\star/dr=f(r)^{-1}$. Choosing an appropriate integration constant, we obtain
\begin{equation}
    r^\star = \frac{3}{\Lambda}\left(\frac{r_B\log\left|\frac{r-r_B}{2M}\right|}{(r_C-r_B)(2r_B+r_C)}-\frac{r_C\log\left|\frac{r_C-r}{2M}\right|}{(r_C-r_B)(r_B+2r_C)}+\frac{(r_B+r_C)\log\left|\frac{r+r_B+r_C}{2M}\right|}{(r_B+2r_C)(2r_B+r_C)}\right).
\end{equation}
The metric in $(u,r,\theta,\phi)$ coordinates is
\begin{equation}
    ds^2=-f(r)du^2-2du\, dr+r^2(d\theta^2+\sin^2\theta\, d\phi^2).
\end{equation}

We will be evolving null geodesics backwards in time from a point $p$ at future null infinity, as described above. Without loss of generality, we may assume that $p$ satisfies $u=0$, $\theta=0$. Moreover, by spherical symmetry we may assume that each geodesic lies entirely on the union of the half-regions with $\phi=\phi_0$ and $\phi=\phi_0+\pi$, for some $\phi_0$.\footnote{Strictly speaking, the polar coordinate $\theta$ has the domain $[0,\pi]$, but we will sometimes refer to negative $\theta$ by identifying $(-\theta,\phi_0)\sim (\theta,\phi_0+\pi \mod 2\pi)$.} Geodesics with different $\phi_0$ are related by the isometry $\partial/\partial \phi$. Hence, we are looking for null geodesics which satisfy
\begin{equation}
    u \to 0,\; \theta\to 0 \quad\text{as}\quad r\to \infty,\qquad\text{and}\qquad \dot{\phi}=0,
\end{equation}
where a dot represents a derivative with respect to a parameter $\lambda$, which we assume to be affine. The null geodesic equations can be derived from the Lagrangian $L=g_{ab}\dot{x}^a\dot{x}^b$, and we find
\begin{align}
    f(r)\dot{u}+\dot{r} = f(r)\dot{t} &= \text{const.} \equiv E,\\
    r^2\dot{\theta} &= \text{const.} \equiv h,\\
    \dot{r} &= \pm\sqrt{E^2-\frac{h^2f(r)}{r^2}} \label{eq:radial_Lagrangian}
\end{align}
where $E$ and $h$ are constants of motion. For null geodesics these do not have any intrinsic physical meaning, due to the freedom in choosing a new affine parameter, but the impact parameter $q=h/|E|$ does. Unlike in the Schwarzschild spacetime, $E$ need not be positive; in particular, the geodesic that passes through the bifurcation sphere of the cosmological horizon (in the black hole frame) has $E=0$, and the geodesics that pass to its right (through the upper half of $\mathcal{H}_C^-$ in Figure \ref{fig:penrose}) have negative $E$. We can treat each case independently, and set $E=-1,0,1$ as appropriate and vary $h$.

While these equations can be solved analytically \cite{Hackmann:2008zz}, the solutions are cumbersome, and it is easier to evolve them numerically. However, the square root in the radial equation makes the numerical implementation messy and prone to errors in the vicinity of a turning point. Thus, we instead use the Hamiltonian formalism with Hamiltonian
\begin{equation}
    H=\frac{1}{2}g^{ab}p_ap_b = \frac{f(r)p_r^2}{2}-p_rp_u+\frac{p_\theta^2}{2r^2}.
\end{equation}
Note that $p_u=-E$ and $p_\theta=h$. Hamilton's equations now read
\begin{align}
    \dot{u} &= -p_r \label{eq:u_Ham} \\ 
    \dot{r} &= f(r)p_r+\sgn(E)\\
    \dot{\theta} &= \frac{q}{r^2}\label{eq:theta_Ham} \\ 
    \dot{p}_r &=-\frac{1}{2}p_r^2\partial_r f(r)+\frac{q^2}{r^3}, \label{eq:pr_Ham} 
\end{align}
where $\sgn(E)$ is the sign of $E$.

We are interested in studying the geometry of horizon cross-sections as a function of time. Unfortunately, the coordinate $t$ is not a suitable time function, as it diverges on $\mathcal{H}_C^+$. We require a time function $T$ that is well-defined for $r_B<r<\infty$. Furthermore, it is desirable that $T=\text{const.}$ hypersurfaces tend to $t=-\infty$ as $T\to-\infty$ in the region where both are defined, and that every generator crosses at least one constant-$T$ hypersurface. A simple function that achieves this is given by
\begin{equation}\label{eq:T_def}
    T=t+\frac{3}{\Lambda}\frac{r_C}{(r_C-r_B)(r_B+2r_C)}\log\left|\frac{r_C-r}{2M}\right|. 
\end{equation}
It is easy to show that all constant-$T$ hypersurfaces are spacelike. $T$ satisfies the equation
\begin{equation}
    \dot{T} = \dot{t}-\frac{3}{\Lambda}\frac{r_C}{(r_C-r_B)(r_B+2r_C)(r_C-r)}\dot{r},
\end{equation}
and one can easily find the geodesic equations parametrised by $T$ with $dx^a/dT=\dot{x}^a/\dot{T}$.

In theory, one should evolve these equations from $p$ backwards in time, i.e.~starting from $T=\infty$. This is, of course, impossible to do numerically, so instead we use a very large initial parameter, and use the asymptotic expansions of $r$, $p_r$, $u$ and $\theta$ evaluated at the initial parameter to set our initial conditions. (This method was already employed in \cite{Emparan:2017vyp,Gadioux:2024tlm}, for example.) It turns out to be easier to perform the asymptotic expansions in $\lambda$ than in $T$, and thus we write our initial conditions as $r(T(\lambda))=r(\lambda)$ for some large $\lambda$, and similarly for $p_r$, $u$ and $\theta$. We use \eqref{eq:radial_Lagrangian} to compute the expansion of $r(\lambda)$, and then substitute into \eqref{eq:u_Ham}, \eqref{eq:theta_Ham}, \eqref{eq:pr_Ham} and \eqref{eq:T_def} to find the other expansions. We obtain, to second order in $1/\lambda$,
\begin{align}
    r(\lambda) &\sim Q\, \lambda + \frac{q^2}{2Q^3\lambda}-\frac{q^2M}{2Q^4\lambda^2}+\mathcal{O}(\lambda^{-3}) \\
    p_r(\lambda) &\sim \frac{3\left(\sgn(E)-Q\right)}{\Lambda\, Q^2\lambda^2}+\mathcal{O}(\lambda^{-4})\\
    u(\lambda) &\sim \frac{3\left(\sgn(E)-Q\right)}{\Lambda\, Q^2\lambda}+\mathcal{O}(\lambda^{-3})\\
    \theta(\lambda) &\sim -\frac{q}{Q^2\lambda} + \mathcal{O}(\lambda^{-3})\\
    T(\lambda) &\sim \frac{3}{\Lambda}\left[\frac{r_C\, \log(Q\lambda/2M)}{(r_C-r_B)(r_B+2r_C)}+\frac{\sgn(E)}{Q^2\lambda}-\frac{r_C^2}{(r_C-r_B)(r_B+2r_C)Q\lambda}\right.\nonumber \\
    &\qquad \left.+\frac{r_C(q^2-Q^2r_C^2)}{2(r_C-r_B)(r_B+2r_C)Q^4\lambda^2}\right] + \mathcal{O}(\lambda^{-3}),
\end{align}
where for clarity we introduced $Q=\sqrt{1+q^2\Lambda/3}$. The constant term in $r(\lambda)$ was set to zero by employing the freedom to shift $\lambda$.

We are now in a position to evaluate the geodesics numerically, but recall that to find the event horizon, we need to determine the past endpoint along each generator, if it exists. In general this is a difficult problem, but the isometry $\partial/\partial \phi$ greatly simplifies it in our case. It is clear that generators with equal and opposite impact parameter will intersect, by symmetry, at $\theta=\pi$. Hence, we need only integrate the geodesic equations until that point, or until the geodesic reaches $\mathcal{H}_B^-$ or $\mathcal{H}_C^-$, whichever comes first.\footnote{We have not found any evidence that generators with $|q_1|\neq|q_2|$ can intersect before reaching $\theta=\pi$.} In the latter case, $u$ will diverge as the curve approaches $\mathcal{H}_C^-$; to avoid any issues, we switch to advanced coordinates $(v\equiv t+r^\star,r,\theta,\phi=\phi_0)$ when a turning point in $r$ is reached. (Note that a geodesic from infinity can reach at most one turning point in $r$ before returning to infinity.)

We have implemented the null geodesic equations in Mathematica and solved them with the \texttt{NDSolve} function. We usually set our initial conditions at $\lambda=10^5$, although it was sometimes useful to use larger values when calculating late-time quantities, e.g.~the final area of the horizon. We found very little dependence in the accuracy of the numerical integration on the value of $\lambda$ used to set initial conditions, over several orders of magnitude of $\lambda$. We estimate that the coordinates of the caustic points are accurate to within seven significant figures. This estimate is based on a comparison of the numerics with the analytical solution to the $\theta$ equation, which takes a reasonably simple form in an appropriate range of $q$ (see Appendix \ref{app:error_analysis}). Our numerical approach is limited in the range of $M\sqrt{\Lambda}$ it can probe, as for values that are very close to the endpoints of the domain $M\sqrt{\Lambda}\in (0,1/3)$, extremely large quantities are multiplied by extremely small factors, and significant uncertainties can result from those computations. Nevertheless, we are able to probe a sufficiently large subset of the domain, and near enough to these bounds, to be confident about our conclusions.

\section{Results}\label{sec:results}

\subsection{Horizon cross-sections}

\begin{figure}[t]
    \centering
    \includegraphics[width=\textwidth]{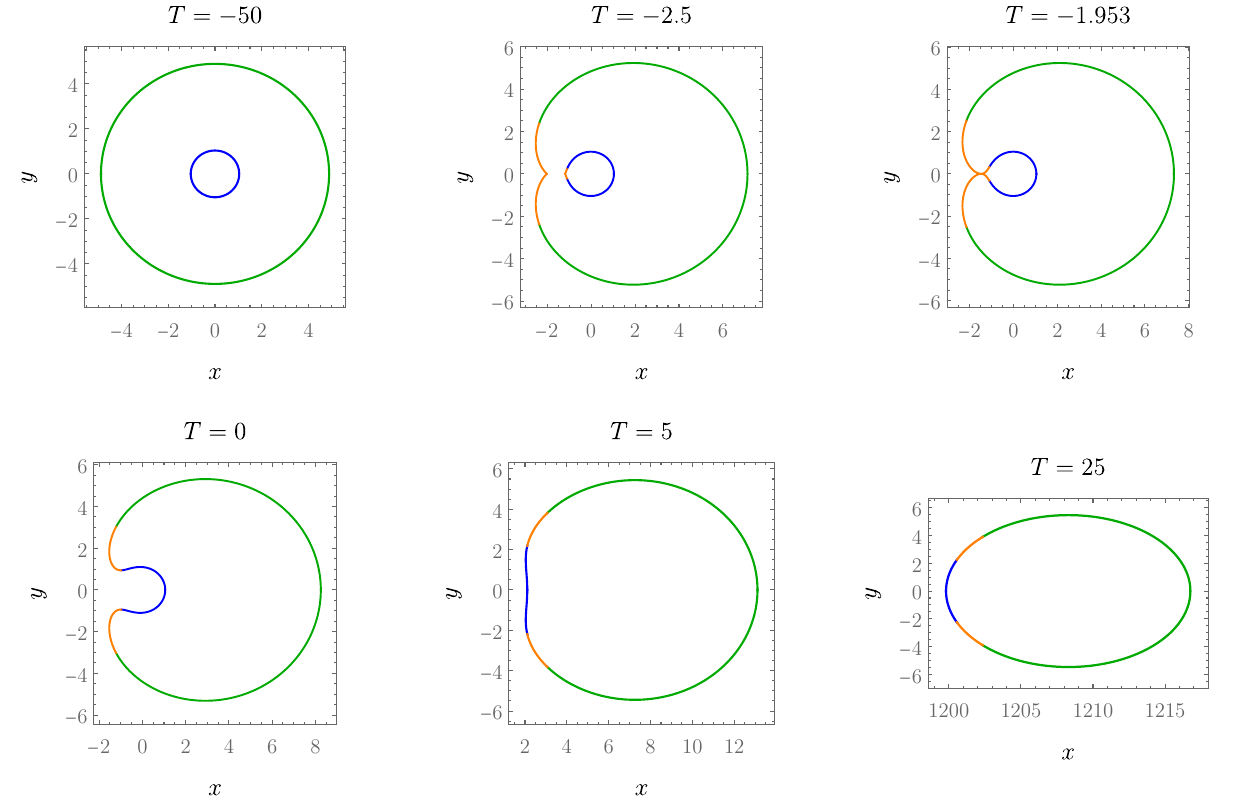}
    \caption{Evolution of the event horizon with $\Lambda=0.1$, $M=0.5$. At early times, there are two nearly-spherical disconnected horizon cross-sections. These develop increasingly-stretched and pointed tips which approach each other. The tips are caustic points. At $T\approx-1.953$ the tips touch: this is the point of merger in this foliation. At later times, the now connected horizon becomes convex and smooths out into a spherical shape. (It looks ellipsoidal here as a result of the choice of coordinates.) Blue segments are points on black hole generators (originating from the white hole horizon $\mathcal{H}_B^-$), orange segments are points on caustic generators, and green segments are points on cosmological horizon generators (coming from $\mathcal{H}_C^-$).}
    \label{fig:merger_evolution}
\end{figure}

The evolution of the event horizon can be studied by plotting cross-sections on constant-$T$ surfaces. Due to the $\partial/\partial \phi$ symmetry, it is sufficient to plot the $(r,\theta)$ part; the full $2$-dimensional cross-sections can be recovered by revolving around the $\theta=0$ axis. It is useful for plotting purposes to introduce Cartesian-like coordinates,
\begin{equation}
    x=r \cos \theta,\qquad y=r \sin \theta,
\end{equation}
with the axis of revolution now being the $x$ axis. An example of an evolution is shown in Figure \ref{fig:merger_evolution} for a cosmological constant $\Lambda=0.1$ and black hole mass $M=0.5$ ($M\sqrt{\Lambda}\approx 0.158$). Initially, for $T\to-\infty$, we observe two disconnected spherical horizon cross-sections, representing the black hole and cosmological horizon in the observer frame, respectively.\footnote{Whenever we mention a ``horizon'' at a particular moment of time, we are really referring to a horizon cross-section, but for conciseness we will often drop ``cross-section''.} The black hole then gradually approaches the cosmological horizon along the negative $x$ direction. During this process, each horizon cross-section develops a sharp tip protruding towards one another. The tips are caustic points. In the full time evolution, these points form a single spacelike caustic line at $y=0$. At some time $T=T_\star$ the two tips touch: this is the point of merger. We emphasise that the point of merger depends on the foliation; however, it will always occur at a caustic point, since it must be at a past endpoint, and all past endpoints are caustic points in our system. For $T>T_\star$, the horizon cross-section is connected and smooth, i.e.~it no longer contains caustic points. Eventually, it becomes convex, and asymptotes to a spherical shape as $T\to\infty$. The process is qualitatively similar for all $0<M\sqrt{\Lambda}<1/3$, except that the initial size of the black hole (with respect to the cosmological horizon) decreases with $M\sqrt{\Lambda}$. With our choice of foliation, the time of merger $T_\star$ decreases with increasing $M\sqrt{\Lambda}$.

Many of the above features are reminiscent of those observed by Emparan and Mart\'inez in the merger of two Schwarzschild black holes in the infinite mass ratio limit \cite{Emparan:2016ylg}. The main difference is that the cosmological horizon, which replaces the large planar black hole, has closed cross-sections of finite size. In fact, we will argue in Section \ref{sec:small_Lambda} that the Schwarzschild merger is the limit of zero cosmological constant of the Schwarzschild-de Sitter system we are considering here.

A potential quantity of interest is a measure of the duration of the merger. However, it is unclear to us how to define one that is physically meaningful, since the spacetime point of merger depends on the foliation used. For simplicity, we will use the foliation defined by $T$ in \eqref{eq:T_def}.

\begin{figure}
    \centering
    \includegraphics[width=0.5\textwidth]{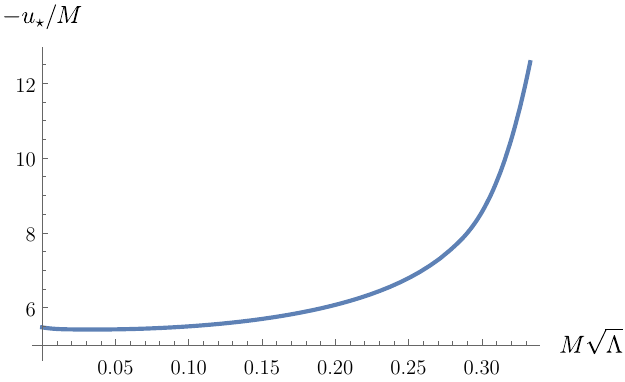}
    \caption{Duration of the merger, characterised by $-u_\star$ (in units of $M$), as a function of $M\sqrt{\Lambda}$. The duration reaches a minimum of approximately $5.41$ at $M\sqrt{\Lambda}\approx 0.034$. At $M\sqrt{\Lambda}=0$, the duration is $5.47$, while a maximum of $12.62$ is reached when $M\sqrt{\Lambda}=1/3$.}
    \label{fig:merger_duration}
\end{figure}

As seen in Figure \ref{fig:merger_evolution}, just after merger, the horizon cross-section on the $xy$ plane appears non-convex.\footnote{The induced metric on the $xy$ plane is not Cartesian, so the concept of convexity is not as straightforward as in Euclidean geometry. However, we are only interested in a heuristic definition of the duration of merger, so we will ignore this detail for this discussion.} As the black hole is progressively absorbed by the cosmological horizon, the horizon becomes closer to convexity. Due to the reflection symmetry of the horizon in the $x$ axis, it is clear that the final point of non-convexity must be on $y=0$, and it lies on a black hole generator. In fact, it lies on the generator with constant $u\equiv 0$. We define the duration of the merger, or perhaps more precisely its relaxation time, as the difference in the retarded times of the point of merger and the last point of non-convexity. Given the above comments, this reduces to $-u_\star$, where $u_\star$ is the retarded time at the point of merger. This is equivalent to the definition used in \cite{Emparan:2016ylg}, although the relation to convexity is not clear in their case as this condition is only reached at future infinity. We note, however, that our time functions differ, so we do not expect to obtain their result in the limit of zero cosmological constant. With $-u_\star$ in units of the black hole mass $M$, we find, curiously, a non-monotonic relationship as a function of the dimensionless parameter $M\sqrt{\Lambda}$. As shown in Figure \ref{fig:merger_duration}, a minimum of $5.41$ is reached at $M\sqrt{\Lambda}\approx 0.034$. The maximum, $12.62$, is reached in the Nariai limit $M\sqrt{\Lambda}\to 1/3$, whereas the duration of merger approaches $5.47$ when $M\sqrt{\Lambda}\to0$. We remark that if we were to express $-u_\star$ in units of $1/\sqrt{\Lambda}$, the duration of merger would be a monotonically-increasing function which vanishes as the black hole is shrunk to zero mass, $M\sqrt{\Lambda}\to0$.

\subsection{Area Increase}\label{sec:area}

The main advantage of the Schwarzschild-de Sitter merger over the asymptotically-flat Schwarzschild merger in \cite{Emparan:2016ylg} is that the horizon cross-sections are of finite size. This implies that we can study the area, and therefore the entropy, of the total system, instead of having to restrict to a subset of the generators. In particular, we can determine the total area growth, as well as compare the contributions to this area increase from black hole, cosmological horizon and caustic generators. This is not possible in the Schwarzschild merger, since the large black hole is infinitely large on all time slices.

Taking advantage of the axisymmetry of the horizon cross-sections, and parametrising $r$ as a function of $\theta$, the area of a constant-$T$ cross-section of the event horizon is given by
\begin{equation}\label{eq:area_function}
    A(T) = 2\pi\int d\theta\, r\sin \theta \sqrt{\left[f(r)^{-1}-f(r)\left(\frac{3}{\Lambda}\frac{r_C}{(r_C-r_B)(r_B+2r_C)(r_C-r)}\right)^2\right]\left(\frac{dr}{d\theta}\right)^2+r^2}.
\end{equation}
Note that the function $r(\theta)$ is double-valued, and thus one must take particular care to include both branches. Before merger (see the first two plots of Figure \ref{fig:merger_evolution}), the integral splits into two pieces: one for each of the disconnected pieces of the horizon cross-section. The integral is taken from $0$ to $\pi$ in each case. After merger, the horizon cross-section is connected but $r(\theta)$ is still double-valued (e.g.~plots 4 to 6 of Figure \ref{fig:merger_evolution}; recall that $r$ is measured from the origin $(0,0)$). In this case, the integral must be split at the maximum angle $\theta_0$ that the horizon cross-section reaches, and the limits are then $0$ to $\theta_0$ for both branches. If we are only interested in the area for a subset of generators, then the integral should be taken over the appropriate range of $\theta$, and the correct branch must be chosen, if applicable.

Let us denote the initial area by $A(-\infty)\equiv A(T\to-\infty)$. The horizon cross-section with $T=-\infty$ is composed of two disconnected pieces, $\mathcal{H}_B^-\cap\mathcal{H}^+$ and $\mathcal{H}_C^-\cap\mathcal{H}^+$, representing the black hole and cosmological horizons, respectively. As explained in Section 2.1 of \cite{Gadioux:2024tlm}, despite being non-smooth, $\mathcal{H}_B^-\cap\mathcal{H}^+$ is isometric to a cross-section of $\mathcal{H}_B^+$, and thus has an area of $4\pi r_B^2$. A similar reasoning applies to $\mathcal{H}_C^-\cap\mathcal{H}^+$. Moreover, the caustic generators do not contribute to the area in the limit $T\to-\infty$. Hence, we have
\begin{equation}\label{eq:A_minf}
    A(-\infty)=4\pi(r_B^2+r_C^2).
\end{equation}
At very late times, the black hole has been absorbed, and thus the observer at $p$ only sees an empty de Sitter universe. The final area is therefore that of a cosmological horizon in a pure de Sitter spacetime, and is given by
\begin{equation}\label{eq:A_pinf}
    A(\infty)=4\pi(3/\Lambda).
\end{equation}
These exact results provide useful consistency checks for the numerics. Note that attempting to compute the area numerically at very large $T$ can lead to larger errors than at slightly smaller $T$, since the horizon cross-section moves away from the origin of the coordinate system, so any error in the angular coordinate is magnified by large radial factors. Thankfully, the area plateaus relatively quickly, so one can obtain accurate results with sufficiently small values of $T$.

One small remark is in order before we present our results. We have two lengthscales in our problem, $M$ and $\sqrt{\Lambda}$, but since we can construct a dimensionless quantity $M\sqrt{\Lambda}$ out of them, one of them is redundant. We can write the area in units of $M^2$ or in units of $1/\Lambda$. Mathematically these are equivalent, but the interpretation is different. If we use $M^2$ as our unit of area, we are effectively fixing the mass of the black hole and varying the cosmological constant. We prefer to instead use units of $1/\Lambda$, since the cosmological constant is a fundamental constant of nature, and so should be fixed.

\begin{figure}
    \centering
    \includegraphics[width=0.6\textwidth]{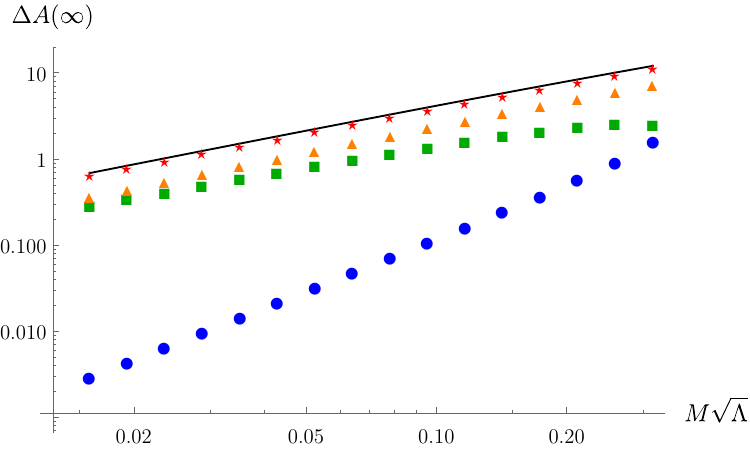}
    \caption{Area increase $\Delta A(\infty)$, in units of $1/\Lambda$, as a function of $M\sqrt{\Lambda}$. The area increase from black hole, cosmological horizon and caustic generators are plotted with blue discs, green squares and orange triangles, respectively. The red stars give the total area increase determined numerically; this is in good agreement with the analytical result (black line). Note that the axes are logarithmic.}
    \label{fig:total_area_Lambda}
\end{figure}

We may define the \emph{area increase} at a time $T$ as
\begin{equation}
    \Delta A(T)\equiv A(T)-A(-\infty).
\end{equation}
We can apply this definition for the entire set of generators, or for a subset of them, e.g.~only the caustic generators, or only the black hole generators, etc. Let us first compute this quantity as a function of $M\sqrt{\Lambda}$ for $T=\infty$, and also study how it is distributed between black hole, cosmological horizon and caustic generators. Our results are shown in Figure \ref{fig:total_area_Lambda}.

Since we are expressing the area in units of $1/\Lambda$, and thus are varying the size of the black hole, we expect the total area increase to be larger when the black hole has larger mass, i.e.~we expect a monotonically-increasing function of $M\sqrt{\Lambda}$. This is indeed what we observe (red stars and black line in Figure \ref{fig:total_area_Lambda}; the black line is the analytical result $A(\infty)-A(-\infty)$ obtained from \eqref{eq:A_minf} and \eqref{eq:A_pinf}). Note that if we had used units of $M^2$ for the area, we would have obtained a monotonically-\emph{decreasing} area change. This is because the initial area of the event horizon (in units of $M^2$) is larger for smaller values of $M\sqrt{\Lambda}$.\footnote{That the initial area is larger does not guarantee that the area growth is also larger, of course, but it turns out that the ``average expansion'' along the generators decreases sufficiently slowly, as a function of $M\sqrt{\Lambda}$, that this relationship is indeed true. We will comment further on this later.} Moreover, we find that the cosmological horizon generators (green squares) always provide a larger contribution to the area growth than the black hole generators (blue discs), except in the Nariai limit (i.e.~$M\sqrt{\Lambda}\to1/3$), where the two horizons coincide and the contributions equalise. Again, the main explanation for this result is the relative initial areas of the black hole and the cosmological horizon: the cosmological horizon is initially larger than the black hole, and thus undergoes more growth. Perhaps more interesting is that the caustic generators (orange triangles) always provide the dominant contribution, for any $M\sqrt{\Lambda}$. In particular, this continues to hold true in the limit $M\sqrt{\Lambda}\to 0$, when we might expect cosmological horizon generators to dominate the area growth, as the black hole is much smaller than the cosmological horizon.

The total area increase lies on a nearly straight line on the logarithmic plot. This can be seen clearly in the limit of small mass, $M\sqrt{\Lambda}\ll 1$. The total change in area is given by $\Delta A(\infty)=12\pi/\Lambda-4\pi(r_B^2+r_C^2)$. To find how $r_B$ and $r_C$ vary with $M$, we can write the (real) roots of the cubic polynomial $r f(r)$ in terms of trigonometric functions,
\begin{equation}
    r=\frac{2}{\sqrt{\Lambda}}\cos\left(\frac{\arccos\left({-3M\sqrt{\Lambda}}\right)+2\pi k}{3}\right),\qquad k=0,1,2.
\end{equation}
Expanding, we find that for small $M\sqrt{\Lambda}$, $r_B$ corresponds to $k=2$ and $r_C$ corresponds to $k=0$. We have
\begin{equation}
    r_B = 2M + \mathcal{O}\left(M^3\Lambda^{3/2}\right),\qquad
    r_C = \sqrt{\frac{3}{\Lambda}}-M-\frac{M^2\sqrt{3\Lambda}}{2} + \mathcal{O}\left(M^3\Lambda^{3/2}\right),
\end{equation}
and thus,
\begin{equation}\label{eq:total_area_increase}
    \Delta A(\infty) = \frac{8\sqrt{3}\pi M}{\sqrt{\Lambda}} - 8\pi M^2 + \mathcal{O}\left(M^3\Lambda^{3/2}\right).
\end{equation}
We see that the area growth increases linearly with $M\sqrt{\Lambda}$. It is interesting to note that, while the above analysis is only valid for small $M\sqrt{\Lambda}$, this behaviour persists to good accuracy along the entire domain $M\sqrt{\Lambda}\in (0,1/3)$. Indeed, the true relationship (the black line in Figure \ref{fig:total_area_Lambda}), grows just slightly more slowly than linearly for larger $M\sqrt{\Lambda}$.

\begin{figure}
    \centering
    \includegraphics[width=0.6\textwidth]{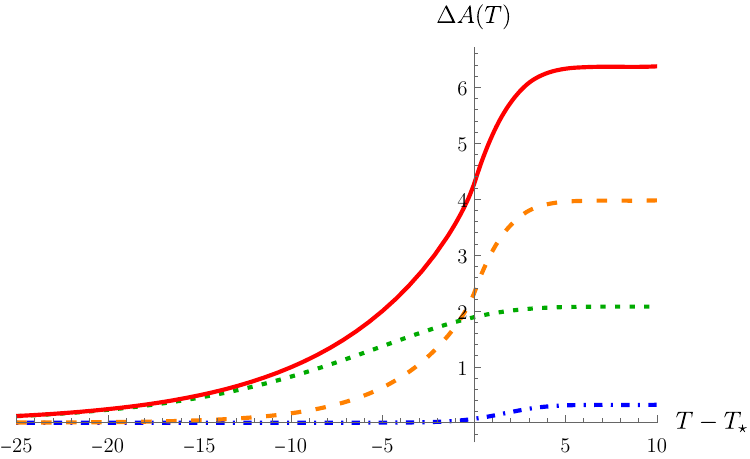}
    \caption{Area increase $\Delta A(T)$ for the merger with $\Lambda=0.1$, $M=0.5$. Contributions from black hole, cosmological horizon and caustic generators are shown in dash-dotted blue, dotted green, and dashed orange, respectively. The solid red curve gives the sum of these. For smaller $M\sqrt{\Lambda}$, the curves are qualitatively similar, but the increase is less sudden.}
    \label{fig:area_evolution}
\end{figure}

Now fixing $M\sqrt{\Lambda}$, we can study how the change in area, $\Delta A(T)$, varies as a function of time. An example is shown in Figure \ref{fig:area_evolution} for a merger with $\Lambda=0.1$ and $M=0.5$. We find that the area plateaus quite rapidly after merger, and this holds true for all $M\sqrt{\Lambda}$ that we have tested. The greatest rate of increase occurs near the time of merger $T=T_\star$, and the dominant contribution to this large rate is from the caustic generators. This is not surprising, since in addition to the expansion of pre-existing caustic generators, there is also the addition of new caustic generators at every time $T\leq T_\star$. Moreover, we expect, intuitively, that most of the distortions that the event horizon experiences will take place near the point of merger, where the caustic generators accumulate. Interestingly, the maximum rate of increase from black hole generators occurs after merger, whereas that from the cosmological horizon generators occurs before merger.

The evolution of $\Delta A(T)$ illustrated in Figure \ref{fig:area_evolution} is qualitatively similar for all values of the black hole mass; the main difference is that for smaller $M\sqrt{\Lambda}$, the increase is less sharp, and spread more evenly in $T\in(-\infty,T_\star)$. This may perhaps be explained by the fact that for small $M\sqrt{\Lambda}$, the distance separating the black hole to the cosmological horizon is larger. Thus, the approach takes longer---this is seen in how $T_\star$ varies with $M\sqrt{\Lambda}$---and the area can increase more steadily. To quantify this effect, we measure how long before merger the total area increase reaches half of its maximum value, i.e.~we compute $T_{1/2}-T_\star$ as a function of $M\sqrt{\Lambda}$, where $\Delta A(T_{1/2})=\Delta A(\infty)/2$. We find that a power law $T_{1/2}-T_\star \sim a+b(M\sqrt{\Lambda})^{-3/2}$, with $a\approx 1.06$, $b\approx-0.16$, agrees very well with the data. (The data probe the range $M\sqrt{\Lambda}\in [0.019,0.316]$, approximately.)

Up to this point, our discussion mostly ignored how large the initial black hole and cosmological horizons are. When $M\sqrt{\Lambda}$ is small, the cosmological horizon is much larger than the black hole, and thus we should expect cosmological horizon generators to provide a larger contribution to the increase in area. However, perhaps a better measure of how ``active'' or ``important'' a set of generators are in the merger process is how much it expands by. The \emph{relative area increase}, 
\begin{equation}
    \delta A(T)\equiv \Delta A(T)/A(-\infty),
\end{equation}
which factors out the initial area, allows us to make this idea precise. Note that when considering a subset of generators (e.g.~only black hole generators), we must divide by the initial area of the part of the horizon that this subset constitutes, not the area of the entire horizon cross-section.

\begin{figure}
    \centering
    \includegraphics[width=0.6\textwidth]{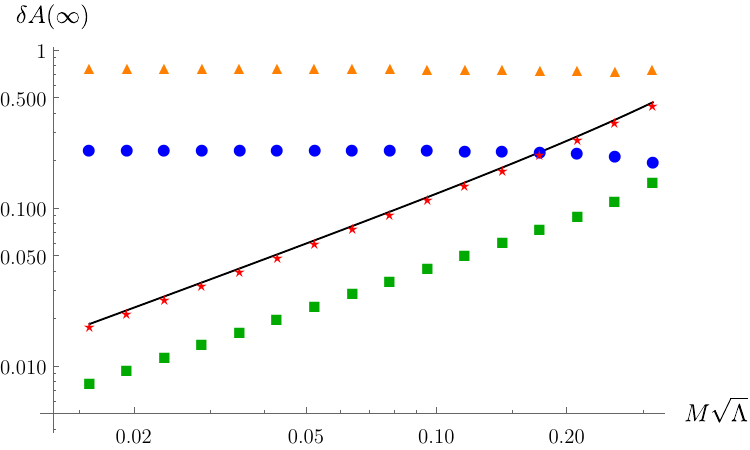}
    \caption{Relative area increase $\delta A(\infty)$ as a function of $M\sqrt{\Lambda}$. The relative increase from black hole generators and cosmological horizon generators are plotted with blue discs and green squares, respectively. The orange triangles represent the black hole generators plus the caustic generators which we associate with the black hole (these are the generators that do not encounter a radial turning point). The relative increase for all generators combined is given by red stars. The black line is the corresponding analytical result.}
    \label{fig:relative_area}
\end{figure}

We find, unlike for the total area increase, that the black hole generators have a larger contribution to $\delta A(\infty)$ than the cosmological horizon generators, for any size of the black hole, although again these contributions tend to the same value as $M\sqrt{\Lambda}\to 1/3$ (see Figure \ref{fig:relative_area}). This implies that the black hole generators experience larger expansion, which we can associate to the more intense dynamics they encounter (the generators of a stationary black hole, which by definition is not dynamical, have zero expansion). The contribution from such generators is largest in the limit of small black hole mass, and decreases with accelerating rate with increasing values of $M\sqrt{\Lambda}$. On the other hand, the relative area increase for both the cosmological horizon and the whole horizon tends to zero as $M\sqrt{\Lambda}\to 0$. This is easily explained from Eq.~\eqref{eq:total_area_increase}: dividing by the initial area $4\pi(r_B^2+r_C^2)$ and expanding, we have
\begin{equation}\label{eq:relative_area_increase}
    \delta A(\infty)=\frac{8\pi M\sqrt{3/\Lambda}-8\pi M^2+\cdots}{4\pi\left(3/\Lambda-2M\sqrt{3/\Lambda}+\cdots\right)} = 2M\sqrt{\frac{\Lambda}{3}}+\cdots,
\end{equation}
where the ellipses denote higher-order terms. A line of best fit of our numerical results gives a slope of approximately $+1.02$, indicating that the linear dependence on the mass remains reasonably accurate even for large values of $M\sqrt{\Lambda}$. The exact result increases slightly faster than linearly as the black hole is made more massive.

We cannot probe the relative area increase from the caustic generators independently, since at $T\to-\infty$ they have zero area. However, we can, following \cite{Emparan:2016ylg}, associate some of the caustic generators to the black hole, and the rest to the cosmological horizon, based on whether the generators encounter a radial turning point. None of the black hole generators encounter such a turning point, so we define the caustic generators with no radial turning point to enter the horizon from the ``black hole side''. Curiously, the relative area increase from this group of generators is nearly constant (more precisely, it has a very slight negative slope) over a large range of the parameter space. For small $M\sqrt{\Lambda}$, the black hole generators also have a slowly-decreasing contribution to the relative area increase; this suggests that, in this regime, the number of caustic generators that enter on the black hole side increases at a similar rate to that of the size of the black hole itself. Near the Nariai limit $M\sqrt{\Lambda}\to 1/3$, the contribution of the black hole generators decreases more rapidly, but this is counteracted by a large increase in the number of caustic generators entering on the black hole side.

\subsubsection{The $\Lambda\to 0$ limit}\label{sec:small_Lambda}

As one takes the cosmological constant $\Lambda$ to zero, keeping the mass fixed, the radius of $\mathcal{H}_C^\pm$, $r_C$, tends to infinity, the black hole radius $r_B$ approaches the Schwarzschild radius, and the metric reduces to the Schwarzschild metric. Naively, one may think that the $\Lambda\to 0$ limit represents an isolated, non-rotating black hole in an asymptotically flat spacetime, since the cosmological horizon is pushed to infinity. However, our late-time conditions on the event horizon are still imposed, in particular that it must asymptote to a null surface at future infinity, and thus a better interpretation is that the $\Lambda\to 0$ limit corresponds to the merger of two Schwarzschild black holes in the infinite mass ratio limit, as originally studied by Emparan and Mart\'inez \cite{Emparan:2016ylg}.

There is a significant issue with the above idea, however. The causal structures of Schwarzschild and Schwarzschild-de Sitter are very different, and the points $p$ on $\mathcal{I}^+$ from which the horizons are defined live in different parts of the Penrose diagram. Indeed, compared with Figure \ref{fig:penrose}, the Penrose diagram for the Schwarzschild merger only consists of the central diamond: future null infinity $\mathcal{I}^+$ has collapsed onto $\mathcal{H}_C^+$, and $\mathcal{H}_C^-$ is replaced by past null infinity $\mathcal{I}^-$. Since this change only occurs at $\Lambda=0$ (for all $\Lambda>0$ the Penrose diagram is as in Figure \ref{fig:penrose}), taking the limit of zero cosmological constant is not a trivial matter.

Thankfully, Geroch provided a solution in a fascinating paper concerning the limits of spacetimes \cite{Geroch:1969ca}. The main issue with taking the limit of spacetimes, say as a parameter $\Lambda$ tends to zero, is that coordinate transformations can lead to very different limiting spacetimes. In order to make the limits well-defined, Geroch constructs a five-dimensional manifold $\mathcal{M}$, with the parameter $\Lambda$ promoted to a scalar field such that slices of constant $\Lambda$ are the four-dimensional spacetimes one wants to take a limit of, each with the corresponding value of $\Lambda$. He then defines a limit space $\mathcal{M}'$ from $\mathcal{M}$ by adding a suitable boundary (i.e.~$\mathcal{M}'$ is a manifold-with-boundary). Finally, to fix the issue of coordinate transformations, he defines a family of orthonormal tetrads attached to a single point $\mathcal{P}(\Lambda)$ of each constant-$\Lambda$ slice. If this family of tetrads has a well-defined limit, then so does the spacetime, at least in a neighbourhood of $\mathcal{P}(0)$. This is Geroch's framework to determine the limits of spacetimes. Note that the same spacetime can have different limits, depending on the curve $\mathcal{P}(\Lambda)$. For example, as the charge parameter in a Reissner-Nordstr\"om spacetime approaches the value of the mass parameter, the limiting spacetime is either the extremal Reissner-Nordstr\"om or the Bertotti-Robinson spacetime \cite{Bengtsson:2014fha}.

Bugden and Paganini showed that the Schwarzschild-de Sitter spacetime reduces, in the sense of Geroch, to pure Schwarzschild in the limit of small $\Lambda$ \cite{Bugden:2018ekg}. Unfortunately, they used the bifurcation sphere of the black hole for $\mathcal{P}(\Lambda)$, and this choice does not allow the regions beyond the cosmological horizon---in particular the region near $\mathcal{{I}^+}$---to survive in the limit. The reason is as follows. Consider a point $\mathcal{Q}$ beyond the cosmological horizon, i.e.~$r>r_C$ at $\mathcal{Q}$. The distance to $\mathcal{Q}$ from the reference point $\mathcal{P}$ (the black hole bifurcation sphere) is the sum of the distance to $r_C$ and the distance from $r_C$ to $\mathcal{Q}$, i.e.~$d(\mathcal{P},\mathcal{Q})=d(\mathcal{P},r_C)+d(r_C,\mathcal{Q})$. In the limit $\Lambda\to0$, the first term on the right hand side blows up. Since the distance from $\mathcal{P}$ to $\mathcal{Q}$ should be even larger, the point $\mathcal{Q}$ cannot exist in the limiting spacetime.

It is clear that the same problem arises for any reference point in $r_B<r<r_C$. However, let us make the coordinate change $x=1/r$. The metric now reads
\begin{equation}
    ds^2=-\left(1-2Mx-\frac{\Lambda}{3}\frac{1}{x^2}\right)du^2+\frac{2}{x^2}du\, dx + \frac{1}{x^2}\left(d\theta^2+\sin^2\theta\, d\phi^2\right).
\end{equation}
As is usually done in the study of causal structure, we will perform a conformal transformation $ds^2\to d\Bar{s}^2=x^2 ds^2$, such that the unphysical metric $d\Bar{s}^2$ is regular at $x=0$:
\begin{equation}
    d\Bar{s}^2=-\left(x^2-2Mx^3-\frac{\Lambda}{3}\right)du^2+2du\, dx + d\theta^2+\sin^2\theta\, d\phi^2.
\end{equation}
The set $\{-\infty<u<\infty,x=0\}$ corresponds to $\mathcal{I}^+$, and point $p$ (see Figure \ref{fig:penrose}) is located at the origin of the coordinate system, $u=x=0$.\footnote{We are ignoring angular coordinates here, as they are not important for the causal structure.} The point $p$ will be our reference point $\mathcal{P}$ in the formalism of Geroch. Note that, unlike in the construction of Bugden and Paganini described above, we do not lose any information in either of the regions of the Penrose diagram of Figure \ref{fig:penrose}, since the map from $r$ to $x$ is bijective, even in the limit $\Lambda\to 0$. Let us now take this limit. First, note that $\mathcal{I}^+$ (which has normal $dx$) becomes null as $\Lambda\to 0$, since
\begin{equation}
    \Bar{g}^{xx}\rvert_{\mathcal{{I}^+}} = x^2-2Mx^3-\frac{\Lambda}{3}\biggr\rvert_{x=0}=-\frac{\Lambda}{3} \xrightarrow{\Lambda\to0} 0,
\end{equation}
as required to retrieve a Schwarzschild spacetime. Moreover, the metric trivially reduces to that of a conformally Schwarzschild spacetime. To understand what happens to the generators, we can examine, for fixed $q$, the coordinate $u$ at which the generator crosses $\mathcal{H}_C^+$, and take the limit $\Lambda \to 0$.\footnote{The definition of the impact parameter $q$ in Schwarzschild-de Sitter agrees with that of the Schwarzschild spacetime in the limit $\Lambda\to 0$.} It turns out that all generators with positive energy converge to $u=0$, whereas the rest (with $E\leq0$) diverge to $u=-\infty$. Thus, the latter generators all converge to the limit generator from spacelike infinity to $p$ along $\mathcal{I}^+$ (see the Penrose diagram in Figure 5 of \cite{Gadioux:2024tlm}). The former set corresponds to the rest of the generators, entering either at a caustic point or at the white hole horizon. Thus, in the limit of zero cosmological constant, we seem to recover the precise construction of a merger of two Schwarzschild black holes with an infinite mass ratio.

% Using that $f(r_C)=0$ to write $\Lambda$ in terms of $r_C$, it is easy to show that the metric becomes
% \begin{equation}
%     ds^2=-f(\Tilde{r})du^2-\frac{2}{(1/r_C-\Tilde{r})^2}du\, d\Tilde{r}+\frac{1}{(1/r_C-\Tilde{r})^2}\left(d\theta^2+\sin^2\theta\, d\phi^2\right),
% \end{equation}
% where
% \begin{equation}
%     f(\Tilde{r})=1-2M\left(\frac{1}{r_C}-\Tilde{r}\right) - \left(\frac{1}{r_C^2}-\frac{2M}{r_C^3}\right)\frac{1}{(1/r_C-\Tilde{r})^2}.
% \end{equation}
% As $r_C\to\infty$, this reduces to
% \begin{equation}
%     ds^2=-\left(1+2M\Tilde{r}\right)du^2-\frac{2}{\Tilde{r}^2}du\, d\Tilde{r}+\frac{1}{\Tilde{r}^2}\left(d\theta^2+\sin^2\theta\, d\phi^2\right),
% \end{equation}
% which is the Schwarzschild metric with the coordinate transformation $r\to \Tilde{r}=-1/r$. Note that in the limit $r_C\to\infty$, points with different radial coordinates $r_B<r_1<r_2<r_C$ tend to different coordinate values, $\Tilde{r}_i\to-1/r_i$, i.e.~we do not lose information in this region by having multiple radii converge to the same coordinate. Since we have not modified the retarded time $u$, in the limit $r_C\to\infty$, points retain the same $u$ coordinate. Therefore, our point $p$ on $\mathcal{I}^+$ remains at $u=0$, as required. Thus, in the limit of zero cosmological constant, we seem to recover the precise construction of a merger of two Schwarzschild black holes with an infinite mass ratio.

Some evidence in favour of this proposition comes from our numerical results. Extrapolating our results in the limit of zero cosmological constant, we obtain a relative area increase $\delta A(\infty)$ of the black hole generators of $0.24174$, which agrees with the value obtained by Emparan and Mart\'inez \cite{Emparan:2016ylg} to the fourth significant figure. Furthermore, if we also include the caustic generators with no radial turning point (corresponding to the orange triangles of Figure \ref{fig:relative_area}), the relative area increase tends to $0.79287$---again a difference of less than 0.1\% with \cite{Emparan:2016ylg}.

\begin{figure}
    \centering
    \includegraphics[width=0.6\textwidth]{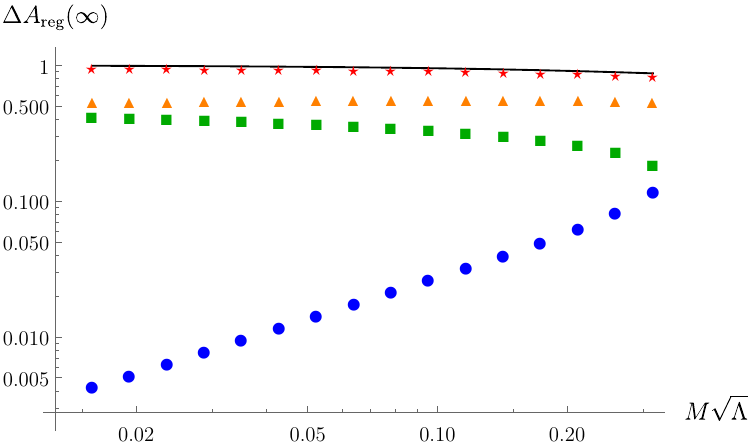}
    \caption{Regularised area increase $\Delta A_{\rm reg}(\infty)$ as a function of $M\sqrt{\Lambda}$. The area increase from black hole, cosmological horizon and caustic generators are plotted with blue discs, green squares and orange triangles, respectively. The red stars give the total regularised area increase determined numerically, while the black line is the analytical result.}
    \label{fig:regularised_area}
\end{figure}

Identifying the $\Lambda\to 0$ limit with the infinite-mass-ratio merger of two non-rotating black holes allows us to regularise infinities present in the original model of \cite{Emparan:2016ylg}. Indeed, the infinite size of the large black hole means that it is impossible to calculate, for example, the total increase $\Delta A(\infty)$ in area. In order to regularise this quantity, we divide it by the factor $8\sqrt{3}\pi M/\sqrt{\Lambda}$, which in the limit of small cosmological constant is the geometric mean of the initial areas of the black hole and the cosmological horizon. Thus, the \emph{regularised area increase} is given by
\begin{equation}
    \Delta A_{\rm reg}(T)\equiv \frac{\Delta A(T)}{8\sqrt{3}\pi M/\sqrt{\Lambda}}.
\end{equation}
We then have, from \eqref{eq:total_area_increase},
\begin{equation}\label{eq:regularised_area}
    \Delta A_{\rm reg}(\infty) = 1-\frac{M\sqrt{\Lambda}}{\sqrt{3}}+\mathcal{O}(M^2\Lambda).
\end{equation}
Thus, we see that in the merger of two Schwarzschild black holes in the infinite mass ratio limit, the regularised area increases by only one unit. We can repeat the same exercise for subsets of the generators; we show our results in Figure \ref{fig:regularised_area}. To estimate the value of $\Delta A_{\rm reg}(\infty)$ when $\Lambda=0$, we fit an interpolating function to our data and extrapolate the limit. We obtain for the cosmological horizon generators, the caustic generators, and all generators, respectively, a regularised area increase of $0.473$, $0.525$ and $0.998$. The latter figure is in very good agreement with \eqref{eq:regularised_area}. Since the black hole generators undergo a finite area growth, their regularised area increase vanishes in the limit of small $\Lambda$.

If we consider a merger of two Schwarzschild black holes of finite masses $M$, $\mu>M$ in a head-on collision, then the area growth, if there were no radiation, would be \cite{Emparan:2016ylg}
\begin{equation}\label{eq:Schw_finite_area}
    \Delta A = 16\pi(M+\mu)^2-16\pi M^2-16\pi \mu^2 = 32\pi M \mu.
\end{equation}
This clearly diverges as $\mu\to\infty$ with $M$ fixed, but we can again regularise the divergence by dividing by the geometric mean of the initial areas, $16\pi M \mu$. This yields the regularised area increase
\begin{equation}
    \Delta A_{\rm reg}=2.
\end{equation}
Naively, the limit $\mu\to\infty$ reduces to the same system as the $\Lambda\to 0$ limit of Schwarzschild-de Sitter, and thus the regularised area increase should agree. However, we instead find that they differ by a factor of 2. It is unclear to us how to interpret this result. Since half of the cosmological horizon generators degenerate to a single limit generator as $\Lambda\to 0$, and since these make up half of the initial area of the cosmological horizon in that limit, perhaps we should regularise by dividing by the geometric mean of the initial black hole area and \emph{half} of the cosmological horizon area. This would account for a factor of $\sqrt{2}$. This still leaves a factor of $\sqrt{2}$ unexplained. Is it related to the energy loss from gravitational waves, which was neglected in \eqref{eq:Schw_finite_area}? This seems unlikely, since the metric is completely static in the infinite mass ratio merger, as well as for all $\Lambda\geq0$ in the Schwarzschild-de Sitter case.\footnote{As noted in \cite{Emparan:2016ylg}, in the extreme mass ratio merger, the radiation zone is pushed away to infinity as $\mu \to \infty$, and furthermore the quasinormal modes are localised, at best, near the photon sphere of the large black hole, i.e.~also at an infinite distance from the small black hole. Thus, there is no source of gravitational waves, and no way for energy to dissipate. The infinite mass ratio merger can be thought of as modelling only the final plunge before merger, and gives the solution near the small black hole to leading order in $M/\mu$. One could then, in principle, add corrections in $M/\mu$ to estimate energy losses, including from a preceding inspiral phase.} Another possibility is that one should replace $M$ by the (non-kinetic) energy of the small black hole in the first term of \eqref{eq:Schw_finite_area}. A body of rest mass $M$ that is stationary at a distance $R$ from a Schwarzschild black hole of mass $\mu$ has energy
\begin{equation}
    E=M\sqrt{1-\frac{2\mu}{R}}.
\end{equation}
To explain the discrepancy of $\sqrt{2}$ in $\Delta A_{\rm reg}$, the energy must satisfy $E=M/\sqrt{2}$, corresponding to $R=4\mu$---this lies between the photon sphere and the innermost stable circular orbit, and is a radius at which strong-gravity effects start to become noticeable. While certainly not a fully satisfactory resolution to the differences in $\Delta A_{\rm reg}$, this may be a plausible heuristic explanation.

\subsection{Conical Singularities}

Before the time of merger, both the black hole and the cosmological horizon develop a conical singularity, with an opening angle that varies in time. The tips of the cones are caustic points. In \cite{Gadioux:2023pmw}, it was shown that in a small neighbourhood of the point of merger, the opening angle $\beta$ scales as $\beta \sim \sqrt{T_\star-T}$.

\begin{figure}
    \centering
    \includegraphics[width=0.6\textwidth]{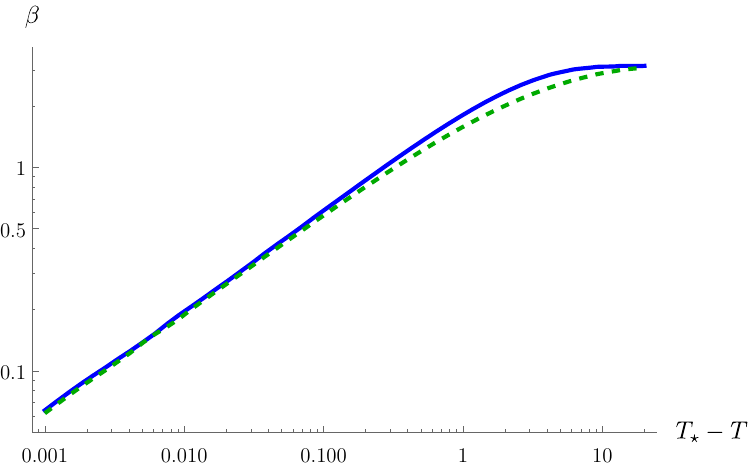}
    \caption{Opening angle of the conical singularities for the merger with $\Lambda=0.1$, $M=0.5$, as a function of time before merger. The solid blue line and the dashed green line correspond to the conical singularities of the black hole and the cosmological horizon, respectively.}
    \label{fig:conical}
\end{figure}

The opening angle can be found from the deficit angle (the angle $\alpha$ that one must cut out from a flat sheet of paper to make a cone) by the formula $\beta=2\arcsin(1-\alpha/2\pi)$. In turn, the deficit angle is $2\pi$ minus the ratio of the circumference of a small circle of constant $\theta$ to its radius, i.e.~$\alpha = 2\pi- 2\pi r/\sqrt{h_{\theta\theta}}$, evaluated at the conical singularity, where $h_{\theta\theta}$ is a component of the induced metric on the horizon cross-section ($h_{\theta\theta}$ is the square-rooted term in \eqref{eq:area_function}). We find that $\beta$ increases monotonically with $T_\star-T$, for both conical singularities. An example is shown in Figure \ref{fig:conical} with $\Lambda=0.1$ and $M=0.5$. At early times, i.e.~large $T_\star-T$, $\beta$ plateaus to $\pi$, as expected, while it vanishes as $T\to T_\star$. There are small differences in the black hole and cosmological horizon curves at early times, but these vanish near the merger, in agreement with the prediction of \cite{Gadioux:2023pmw}, which is universal within a sufficiently small neighbourhood of the point of merger. Including only data with $T_\star-T<0.1$, we obtain (for $M\sqrt{\Lambda}=0.158$) a scaling of $\beta_B\sim(T_\star-T)^{0.51}$ and $\beta_C\sim(T_\star-T)^{0.49}$ for the opening angle of the conical singularities on the black hole and cosmological horizon components, respectively. We observe very little difference for other values of $M\sqrt{\Lambda}$, again corroborating the locally universal nature of black hole mergers.

\section{Conclusions}\label{sec:conclusion}

In \cite{Emparan:2016ylg}, Emparan and Mart\'inez introduced an exactly solvable model of a merger of two Schwarzschild black holes, by taking the mass of one black hole to infinity. This work led to many further examples, derived by simply replacing the small Schwarzschild black hole with another object. In this paper we instead replaced the large black hole with a cosmological horizon in Schwarzschild-de Sitter spacetime.

The qualitative features of the merger are very reminiscent of those of the original model in \cite{Emparan:2016ylg}. On a cross-section at early times, the black hole and the cosmological horizon are disconnected, and each feature a conical singularity, aligned along the axis of merger, and pointing towards each other. These sharp tips are caustic points, which join up in spacetime to form a spacelike line. The point of merger always occurs on this line, regardless of the foliation used. After the merger, the horizon is smooth, and the black hole is gradually absorbed by the cosmological horizon, until the combined horizon eventually relaxes to a spherical shape.

We find that caustic generators are always the dominant source of area growth, for any size of the black hole. In the Nariai limit $M\sqrt{\Lambda}\to 1/3$, where the black hole and cosmological horizons coincide, the contributions to the area increase from the corresponding generators tend to the same value, as expected. For smaller $M\sqrt{\Lambda}$, the cosmological horizon generators induce larger area growth, but this is because the initial surface area is larger; when factoring out the initial area, the contribution from black hole generators becomes more important. This makes sense, since intuitively the black hole is far more involved in the merger process than most of the cosmological horizon.

In the limit of zero cosmological constant, our results for the relative area increase of the black hole generators coincide with that of the original Schwarzschild merger studied in \cite{Emparan:2016ylg}. While the causal structures of the Schwarzschild and Schwarzschild-de Sitter spacetimes differ significantly, we argued, based on the formalism of Geroch \cite{Geroch:1969ca}, that the limit $\Lambda\to 0$ (with $M$ fixed) is well-defined for our construction, and that this limit is precisely the merger of two Schwarzschild black holes in the infinite mass ratio. Using this result, we obtain a regularised area increase for the large black hole, as well as for the whole system. However, it is unclear to us how exactly this should be interpreted.

An obvious extension to this work would be to treat the Kerr-de Sitter spacetime. However, this may be of limited interest, since one can probably form a good guess of the features of this merger simply by examining how the Schwarzschild-de Sitter merger differs from the Schwarzschild case in \cite{Emparan:2016ylg}, and applying this to results for Kerr in \cite{Gadioux:2024tlm}. Nevertheless, perhaps one may find surprising results.

Furthermore, it would be interesting to see whether one can take advantage of Geroch's limit of spacetimes in other situations. For example, divergent quantities appear regularly in physics, especially in the quantum regime; could some of these be regularised by considering a parametrised family of spacetimes, and taking an appropriate limit in some parameter? Another potential application is to introduce a compactification of spacetime depending on some parameter. Then one may try to prove statements on compact manifolds---which may be easier to accomplish than on non-compact ones---and investigate how these behave as the compactification scale is sent to infinity.

It is well-known that horizons have an associated entropy, whether they be a black hole horizon or a cosmological horizon (e.g.~\cite{Gibbons:1977mu}), but their microscopical interpretation remains elusive. That black holes and cosmological horizons can merge suggests that their microstate structures have identical origins, despite black hole horizons being observer-independent, and cosmological horizons not so. Studying the Schwarzschild-de Sitter merger in a semiclassical regime, in a certain limit if necessary (e.g.~$M\sqrt{\Lambda}\to0$, or $M\sqrt{\Lambda}\to1/3$), could perhaps be illuminating. It could also possibly give useful insights into the role that caustics may play. Indeed, it is known that non-smooth entangling surfaces in flat spacetime lead to new subleading terms in the entanglement entropy \cite{Myers:2012vs}. (Some further discussion of this is given in \cite{Gadioux:2023pmw}, although one of its claims is retracted in \cite{Gadioux:2024tlm}.) It would be interesting to investigate entropy and its implications in the Schwarzschild-de Sitter spacetime.

\section*{Acknowledgements}

We thank Harvey Reall for many fruitful discussions throughout this work, and for comments on a draft. MG is supported by an STFC studentship and a Cambridge Trust Vice-Chancellor’s Award. HW received support from the Cambridge Summer Research in Maths (SRIM) programme and St Catharine's College, Cambridge.

\appendix

\section{Error Estimates}\label{app:error_analysis}
To estimate errors on our numerics, we compare the radial position $r$ of caustic points (for an appropriate range of impact parameters $q$) obtained numerically with the analytical result, which we derive below. We make these comparisons over several orders of magnitude in the value of $\lambda$ used to set initial conditions.

To obtain the radial coordinate of the caustic point analytically, we begin by solving for $\theta(r)$, and then invert the function. Let us first write
\begin{align}
    \frac{d\theta}{dr} &= \frac{q}{r\sqrt{r^2-q^2f(r)}}\\
    &= \frac{q}{\sqrt{r}\sqrt{(1+q^2\Lambda/3)r^3-q^2r+2M q^2}}\\
    &= \frac{q}{\sqrt{1+q^2\Lambda/3}}\frac{1}{\sqrt{r(r-r_1)(r-r_2)(r-r_3)}},
\end{align}
where $r_1$, $r_2$ and $r_3$ are the three roots of the cubic expression in the square root of the preceding line. For sufficiently large $q$, these roots are all real and distinct, $r_1>r_2>0>r_3$; we will assume henceforth that we are in this regime. (Note that there is an upper bound on such $q$ that are also caustic points.) We also assume that the caustic has $r>r_1$. Using the condition that $\theta(r=\infty)=0$, we have
\begin{align}
    \theta(r)&=\int_\infty^r\frac{q}{\sqrt{1+q^2\Lambda/3}}\frac{d\Tilde{r}}{\sqrt{\Tilde{r}(\Tilde{r}-r_1)(\Tilde{r}-r_2)(\Tilde{r}-r_3)}}\\
    &=\frac{q}{\sqrt{1+q^2\Lambda/3}}\left[\int_{r_1}^r\frac{d\Tilde{r}}{\sqrt{\Tilde{r}(\Tilde{r}-r_1)(\Tilde{r}-r_2)(\Tilde{r}-r_3)}}-\int_{r_1}^\infty \frac{d\Tilde{r}}{\sqrt{\Tilde{r}(\Tilde{r}-r_1)(\Tilde{r}-r_2)(\Tilde{r}-r_3)}}\right].
\end{align}
This integral can be solved analytically \cite{Gradshtein2007Tois},
\begin{multline}
    \theta(r)=\frac{2q}{\sqrt{1+q^2\Lambda/3}\sqrt{r_1(r_2-r_3)}}\left[F\left(\arcsin\sqrt{\frac{(r_2-r_3)(r-r_1)}{(r_1-r_3)(r-r_2)}}\, \Biggr\rvert\, \frac{r_2(r_1-r_3)}{r_1(r_2-r_3)}\right)\right.\\
    \left.-F\left(\arcsin\sqrt{\frac{(r_2-r_3)}{(r_1-r_3)}}\, \Biggr\rvert\, \frac{r_2(r_1-r_3)}{r_1(r_2-r_3)}\right)\right]
\end{multline}
where $F(\varphi|k)\equiv \int_0^\varphi d\sigma(1-k\sin^2\sigma)^{-1/2}$ is the incomplete elliptical integral of the first kind. Setting $\theta=-\pi$ at the caustic point and inverting, we obtain
\begin{equation}\label{eq:rcaustic}
    r_\text{caustic}(q)=\frac{r_1(r_2-r_3)-r_2(r_1-r_3)\mathcal{S}}{r_2-r_3-(r_1-r_3)\mathcal{S}},
\end{equation}
where
\begin{equation}
    \mathcal{S}=\operatorname{sn}^2\left(F\left(\arcsin\sqrt{\frac{(r_2-r_3)}{(r_1-r_3)}}\, \Biggr\rvert\, \frac{r_2(r_1-r_3)}{r_1(r_2-r_3)}\right)-\frac{\sqrt{1+q^2\Lambda/3}\sqrt{r_1(r_2-r_3)}\pi}{2q}\right)
\end{equation}
and $\operatorname{sn}$ is the Jacobi elliptic sine function.

While not all caustic points are captured by \eqref{eq:rcaustic}---namely, when two of $r_1$, $r_2$ and $r_3$ become complex---we see no reason to believe that the numerical errors on these points should be any higher than for the points we have checked.

\bibliographystyle{JHEP}
\bibliography{paper.bib}

\end{document}